\documentclass[useAMS,usenatbib,referee]{biom}
\def\bSig\mathbf{\Sigma}

\usepackage{setspace}
\usepackage{graphicx}
\usepackage{amsmath}
\usepackage{natbib}
\usepackage{appendix}
\usepackage[section]{placeins}
\usepackage{color}
\usepackage{longtable}
\usepackage{afterpage}
\usepackage{MnSymbol}
\usepackage{rotating}

\title[Truncated Cauchy Combination Test]{Truncated Cauchy Combination Test: a Robust and Powerful P-value Combination Method with Arbitrary Correlations}

\author{Bo Chen$^{1}$, Wei Xu$^{2}$, and Xin Gao$^{3,*}$\email{xingao@yorku.ca} \\
$^{1}$School of Statistics and Data Science, Nankai University, Tianjin, China\\
$^{2}$Department of Statistics, York University, Toronto, Canada \\
$^{3}$Department of Biostatistics, Princess Margaret Cancer Centre, Toronto, Canada}

\begin{document}


\date{{\it Received October} 2007. {\it Revised February} 2008.  {\it
Accepted March} 2008.}



\pagerange{\pageref{firstpage}--\pageref{lastpage}} 
\volume{64}
\pubyear{2008}
\artmonth{December}


\doi{10.1111/j.1541-0420.2005.00454.x}


\label{firstpage}


\begin{abstract}
Cauchy combination test has been widely used for combining correlated p-values, but it may fail to work under certain scenarios. We propose a truncated Cauchy combination test (TCCT) which focus on combining p-values with arbitrary correlations, and demonstrate that our proposed test solves the limitations of Cauchy combination test and always has higher power. We prove that the tail probability of our test statistic is asymptotically Cauchy distributed, so it is computationally effective to achieve the combined p-value using our proposed TCCT. We show by simulation that our proposed test has accurate type I error rates, and maintain high power when Cauchy combination test fails to work. We finally perform application studies to illustrate the usefulness of our proposed test on GWAS and microbiome sequencing data.
\end{abstract}

%



\maketitle


%

\section{Introduction}

Combining test statistics and p-values from multiple tests has been a classical question in statistical testing theory. Standard methods such as Fisher's method \citep{fisher32} and minimal p-value approach \citep{tippett31} have been proved to work well when the tests are independent of each other. However, when there exists unknown or arbitrary dependency between some of the test statistics, the distribution of the combined test statistic or p-value becomes unclear. In such case, permutation tests are often used in practice, but it can be time consuming or even not applicable for achieving extremely small p-values, such as $5 \times 10^{-8}$ or even smaller, which is not uncommon in many application studies, e.g., genome-wide association study (GWAS) \citep{dudbridge08, fadista16}. Besides, it can be also of interest to find the theoretical distribution of the combined test.

To answer this question, \citet{liu20} proposed a Cauchy combination test (CCT), which provided a theoretical answer of the combined p-value with arbitrary correlations of each test. In short, they proposed a combined test statistic. Although its exact distribution is unknown with arbitrary correlations, they showed the tail probability of the combined statistic converges to the tail probability of standard Cauchy distribution as the tail probability goes to zero. This test is very easy and quick to implement in practice, as the combined p-value can be simply obtained from tail probability of standard Cauchy distribution. The method has received a lot of attention in the literature. However, people also found limitations to implement the Cauchy combination test. First, \citet{zhongxue22} showed that the test may fail and have no power at all when combining two perfectly negatively correlated tests. This could happen when combining two one-sided test statistics from the same data but opposite null hypotheses. Next, a more practical situation is that there could be some p-values close or even equal to 1 among all the tests which are being combined. In such case, one test with big p-value close to 1 will make the overall Cauchy statistic insignificant, and thus the test fails to reject the overall null hypothesis even if the rest of the tests being combined have very significant p-values.

In this paper, we propose a truncated Cauchy combination test (TCCT) as an improvement of the original CCT. TCCT solves the above-mentioned limitations, and is theoretically justified to be consistently more powerful than the Cauchy combination test. Similar to the original CCT, we also investigate the tail probability of the distribution of the combined test statistic, and show it converges to Standard Cauchy distribution as the tail probability goes to zero. We recommend TCCT over CCT under all circumstances.

In addition to the classical p-value combination problems, we find that our proposed TCCT can be applied straightforwardly into hypothesis testing questions with longitudinal data. For association study with longitudinal data, functional regression theory has been extensively studied in past years with a significant number of methodology developments and application, and \citet{morris15} gave a comprehensive review. However, compared to functional parameter estimation, methods for hypothesis testing were less comprehensive, especially when the response variable in the regression model is functional. Existing methods for testing the functional parameters in a functional response regression model have a number of limitations, which we will elaborate next.

First, the normality assumption of functional responses is essential in the existing functional response regression models. In practice, it is common that the functional data does not hold the normality assumption, and it is not clear how to estimate and test the functional effects on non-normal functional response variables. Next, functional data are only observed on finite timepoints in practice, and there are usually unequal number of observations for each sample. To deal with the unbalanced data, the common approach is "smoothing first, then estimation" \citep{zhang07}. That is, the first step is to convert the finite discrete data of each sample to continuous function by some smoothing methods. The issue with this two-step approach is that statistical inferences such as test statistics depend on the subjective choice of smoothing strategies. Lastly, due to the arbitrary correlations between different timepoints, the analytical forms of test statistics in functional response regression may not be available, and many testing methods are permutation based or require bootstrapping. The computational cost can be high for large datasets.

Correlations between multiple timepoints can be challenging in functional response regression, but it is not a problem with TCCT, because TCCT needs no assumption on correlations and works with arbitrary correlation structure among timepoints. The implementation of TCCT is simple and straightforward: at each timepoint, no matter all samples or only a subset of samples are observed, we can simply perform an ordinary linear regression test; after that, we can use TCCT to combine each test statistic and achieve the overall test.

The implementation of TCCT on longitudinal data can solve the above-mentioned limitations of existing approaches using functional response regression.  Firstly, the normality assumption is no longer required. For data following other types of distributions at each timepoint, generalized linear model or other appropriate regression model can be used, as long as p-value can be obtained from the test statistic. Secondly, this is a one-step approach and no smoothing method is required, so the subjectivity can be avoided without choosing any smoothing method. Finally, the test statistic has a standard asymptotic distribution, and the computational time can be dramatically improved in most situations. Although CCT may also be implemented to longitudinal data in principle, to our best knowledge, we believe no study up to date has shown how it works for real functional data. In Section 4.2, we use an application study to illustrate the procedure and compare the results for applying both CCT and TCCT to longitudinal data.

In Section 2, we provide all the theoretical details of the proposed truncated Cauchy test and prove the tail probability converges to a standard Cauchy tail. Because the theory only explains situations of small p-values close to 0, we conduct simulation studies in Section 3 to show how the tests actually work at non-trivial type I error rates, and compare TCCT with Cauchy combination test and other classic p-value combination tests. In Section 4, we perform two application studies on GWAS data and a longitudinal data to illustrate the usefulness of our proposed test.

\section{Truncated Cauchy Combination Test (TCCT) Theory}

\subsection{Main theory}
The classical problem of p-value combination is as follows: suppose we have a bunch of p-values from multiple tests, i.e., for  $i=1,2,...,d,$ each test has p-value $p_i$. The correlation between the tests is unknown, and we would like to achieve an overall test statistic based on a combination of these p-values. \citet{liu20} introduced Cauchy combination test (CCT), where the overall test statistic combining each p-value is defined as
$$T_{CCT}=\sum_{i=1}^d w_i \tan\{(0.5-p_i)\pi\},$$
where $w_i$s are nonnegative weights and sum to 1. It is straightforward to see $T_{CCT}$ is dominated by large p-values which are close to 1. If $p_i=1$, then $w_i \tan\{(0.5-p_i)\pi\}= -\infty$, and $T_{CCT}=-\infty$ regardless of the other p-values being combined. To solve this issue, we truncate the lower bound of each term of $w_i \tan\{(0.5-p_i)\pi\}$ at zero and only keep the positive terms in the sum, i.e., $\tan\{(0.5-p_i)\pi\}>0$, or equivalently, $p_i<0.5$. In formal, we define our test statistic as
$$T=\sum_{i=1}^d w_i \tan\{(0.5-p_i)\pi\} I(p_i<0.5),$$
where $I$ is the indicator function. We call this test as Truncated Cauchy combination test (TCCT). We will explain the motivation of the truncation with more details in Section 2.2.

With arbitrary correlations between $p_i$'s, $T$ may not have a closed form distribution. However, we want to show the tail probability of TCCT converges to the tail probability of a standard Cauchy distribution. To show this, we adopt the same bivariate normal distribution assumption from \citet{liu20}, i.e., the p-values $p_i$ come from z-score type of test statistic $X_i$, and each pair of the test statistic follows a bivariate normal distribution. Specifically, let $\boldsymbol{X}=(X_1, X_2,..., X_d)^T$. For any $1 \leq i < j \leq d$, $X_i, X_j \sim N(0,1)$ and $(X_i, X_j)^T$ follows a bivariate normal distribution. It needs to note that this assumption is only for each p-value related test statistic. For the original data being tested, no normality assumption is needed, and TCCT has no restrictions on its distribution type.

Under the bivariate normal condition of each test statistic $X_i$, the overall TCCT statistic can be reformulated as
$$T(\boldsymbol{X})=\sum_{i=1}^d  w_i f(X_i),$$
where $f(x)=h(x)I(h(x)>0)$ and $h(x)=\tan\{[2\Phi(|x|)-3/2]\pi\}$. Clearly function $f(\cdot)$ is function of $h(\cdot)$ truncated at 0. Firstly, we show the relationship of $f(\cdot)$ and $h(\cdot)$ by Lemma 1.

\begin{lemma}
For any $t>0$, $P(f(X_i)>t)=P(h(X_i)>t)$.
\end{lemma}

Following Lemma 1, we have $P(f(X_i) \leq t)=P(h(X_i) \leq t)$, and $P(t_1<f(X_i) \leq t_2)=P(t_1<h(X_i) \leq t_2)$ for any $t_2>t_1>0$. Secondly, we show by Theorem 1 that the tail probability of $T(\boldsymbol{X})$ can be approximated by standard Cauchy distribution.

\begin{theorem}
Under the bivariate normal distribution assumption of $\boldsymbol{X}$, for any fixed $d$, 
$$\underset{t \to +\infty}{\lim} \frac{P(T(\boldsymbol{X})>t)}{P(W_0>t)}=1,$$
where $W_0$ is a standard Cauchy random variable.
\end{theorem}

Note that Theorem 1 assumes the number of tests being combined is finite, which is always the case in practice. But for theoretical interest, we also show that when the number of tests being combined tends to infinity,  $P(T(\boldsymbol{X})>t)$ can still be approximated by standard Cauchy distribution as the tail probability goes to 0. 

\begin{theorem}
Let $d=t^\alpha$ for some $0<\alpha<1/2$. Under the bivariate normal distribution assumption of $\boldsymbol{X}$,
$$\underset{t \to +\infty}{\lim} \frac{P(T(\boldsymbol{X})>t)}{P(W_0>t)}=1,$$
where $W_0$ is a standard Cauchy random variable.
\end{theorem}

We provide the proof of Lemma 1, Theorem 1 and Theorem 2 in Web Appendix A. Following Theorem 1 and 2, it is straightforward to compute the combined p-value using TCCT. For $T(\boldsymbol{X})=t_0$, we have 
$$p_{TCCT}=0.5-(\arctan t_0)/\pi.$$

\subsection{Intuitions behind TCCT}
We have shown by Theorem 1 and 2 that our proposed TCCT has the desired asymptotic property of Cauchy tail probability. However, there may exist other possible ways to improve the original CCT which also solves its limitations, and it may remain unclear to readers why we decide to truncate the original CCT at 0 to define our new test statistic. In section 1, we discussed two scenarios when CCT fails to work. Under both scenarios we are combining some big p-values, which convert to negative values under Cauchy transformation and lead to the overall test statistic of CCT less than or equal to 0. On the contrary, although the Tippett's minimal p-value method may be conservative, it doesn't suffer from the issue of combining big p-values, as only the smallest p-value is considered. With such intuition, our first attempt was to consider choosing only the minimal p-value and transform it to become a test statistic with Cauchy tail. Specifically, let $p_{(1)}=min\{p_1,...,p_d\}$, and $T_{min}=\tan\{(0.5-p_{(1)})\pi\}/d$. We show in Web Appendix B that
$$\underset{t \to +\infty}{\lim} \frac{P(T_{min}>t)}{P(W_0>t)}=1,$$
and thus $T_{min}$ also has Cauchy tail probability and it can also be used as the overall test statistic. The reason we do not suggest using $T_{min}$ is due to its reduced test power. In Web Appendix C, we show that this test has similar performance with Tippett's minimal p-value test, and both tests are relatively conservative in terms of test power. This is because most of the useful information from the other p-values are disregarded.

To achieve higher test power, it is natural to combine the information from p-values of other tests instead of choosing only the smallest p-value. For example, we may consider to combine the smallest $J$ p-values for a predetermined number of $J$. However, the optimal value of $J$ may vary for different data. For example, for 10 tests being combined, it is possible that 2 among 10 have corresponding p-values close or equal to 1, and the optimal approach is to combine the other 8 tests; but for a different set of data, 8 among 10 have corresponding p-values, and the optimal approach is to combine the rest 2 tests. This indicates that there is no optimal value of $J$ which works uniformly best for arbitrary datasets. 

For the above reasons, we consider a truncation at some fixed value for the p-values rather than combining a fixed number of smallest p-values. Specifically, we need to choose a value $p_c$, and disregard all the p-values such that $p_i>p_c$. The rule of thumb to choose the value of $p_c$ is to make the test power on combined p-value as large as possible. When $p_i<0.5$, $\tan\{(0.5-p_i)\pi\}>0$, so including $p_i$ will make the sum $\sum_{i=1}^d w_i \tan\{(0.5-p_i)\pi\}$ larger. Similarly, when $p_i>0.5$, $\tan\{(0.5-p_i)\pi\}<0$, and then excluding $p_i$ will also make the sum larger. Therefore, it is natural to choose $p_c=0.5$, in which case all positive terms in $\sum_{i=1}^d w_i \tan\{(0.5-p_i)\pi\}$ are kept while all negative terms are eliminated, and the sum after truncation, i.e., $\sum_{i=1}^d w_i \tan\{(0.5-p_i)\pi\}I(p_i<p_c)$ is maximized. After the truncation, the number of p-values being combined becomes random, and Theorem 1 and 2 have shown that the test statistic $T$ has the tail probability converging to standard Cauchy distribution.

\section{Simulation Study}
\subsection{Type I error and power evaluation of TCCT}
We have shown in theory that TCCT's tail distribution converges to the tail distribution of a standard Cauchy random variable. In theory, the true type I error of TCCT should converge to the significant level $\alpha$ as $\alpha$ goes to 0. However, the $\alpha$ level of the test is predetermined which cannot be tending to zero in practice, so the true type I error of the TCCT may be different from $\alpha$. In the simulation study, we first estimate the type I error rates under different $\alpha$ levels, and show the empirical type I error rates for TCCT are not far from the targeted values for commonly used $\alpha$s. After that, we assess the test power performance of TCCT under different scenarios. Our primary interest is to compare TCCT over CCT. In addition, we choose Fisher's method and Tippett's minimum p-value method as an example of classic p-value combination methods which require independence assumption. A systematic comparison with more existing p-value combination methods is not the focus of our paper, and \citet{loughin04} had a full discussion on this topic.

We choose $\alpha=0.05, 0.01, 0.001$ and 0.0001, and combine 100 correlated tests. The correlation between each pair of test statistics is set to be 0, 0.3, 0.6 and 0.9. The sample size of each test is $n=100$, and the test is applied to asses the association between a binary covariate and normally distributed response variable under linear regression model. The true effect of the predictor is set as 0 to evaluate type I error, and set as 0.25 to evaluate test power. Table 1 shows the estimated type I error.

\begin{table}[ht]
\centering
\begin{tabular}{rr|rrrr}
  \hline
$\rho$ & $\alpha$ & TCCT & CCT & Fisher & Tippett \\ 
  \hline
0.0 & 0.0500 & 0.07235 & 0.05056 & 0.04941 & 0.05072 \\ 
  0.0 & 0.0100 & 0.01121 & 0.01033 & 0.00993 & 0.01046 \\ 
  0.0 & 0.0010 & 0.00097 & 0.00095 & 0.00098 & 0.00096 \\ 
  0.0 & 0.0001 & 0.00012 & 0.00012 & 0.00016 & 0.00012 \\ \hline
  0.3 & 0.0500 & 0.08315 & 0.06716 & 0.19468 & 0.04184 \\ 
  0.3 & 0.0100 & 0.01421 & 0.01366 & 0.15313 & 0.00979 \\ 
  0.3 & 0.0010 & 0.00108 & 0.00108 & 0.11657 & 0.00090 \\ 
  0.3 & 0.0001 & 0.00013 & 0.00013 & 0.09270 & 0.00013 \\ \hline
  0.6 & 0.0500 & 0.07392 & 0.07045 & 0.25732 & 0.02406 \\ 
  0.6 & 0.0100 & 0.01405 & 0.01399 & 0.22736 & 0.00580 \\ 
  0.6 & 0.0010 & 0.00134 & 0.00134 & 0.19785 & 0.00079 \\ 
  0.6 & 0.0001 & 0.00013 & 0.00013 & 0.17547 & 0.00010 \\ \hline
  0.9 & 0.0500 & 0.05359 & 0.05359 & 0.29549 & 0.00517 \\ 
  0.9 & 0.0100 & 0.01053 & 0.01053 & 0.27177 & 0.00129 \\ 
  0.9 & 0.0010 & 0.00106 & 0.00106 & 0.24641 & 0.00012 \\ 
  0.9 & 0.0001 & 0.00010 & 0.00010 & 0.22674 & 0.00003 \\ \hline
   \hline
\end{tabular}
\vspace{1cm}
\caption{Type I error performance based on 100,000 replications, $d=100$}
\end{table}

We notice that the type I error rates can be slightly inflated for TCCT when $\alpha=0.05$. Similarly, CCT has the same issue, and the inflated type I error is bounded at 0.08 in the worst case. Besides, in accordance with the theory that its tail probability converges to that of standard Cauchy distribution, the type I errors are more accurate for smaller $\alpha$. In comparison, the estimated type I errors from Fisher's method and Tippett's minimal p-value method can be very inaccurate under all $\alpha$ levels when the tests are correlated. The estimated type I error are significantly inflated with Fisher's method and significantly deflated with Tippett's minimal p-value method. We conclude that TCCT and CCT have overall more accurate type I errors.

The power performances of the tests are provided in Table 2. TCCT always have equal or slightly improved test power compared to CCT. Fisher's method and Tippett's minimal p-value method have inflated and deflated power due to their inaccurate type I errors.

\begin{table}[ht]
\centering
\begin{tabular}{rr|rrrr}
  \hline
$\rho$ & $\alpha$ & TCCT & CCT & Fisher & Tippett \\ 
  \hline
0.0 & 0.0500 & 1.0000 & 1.0000 & 1.0000 & 1.0000 \\ 
  0.0 & 0.0100 & 1.0000 & 1.0000 & 1.0000 & 0.9988 \\ 
  0.0 & 0.0010 & 0.9880 & 0.9880 & 1.0000 & 0.8658 \\ 
  0.0 & 0.0001 & 0.5366 & 0.5366 & 1.0000 & 0.4150 \\ \hline
  0.3 & 0.0500 & 0.9930 & 0.9914 & 0.9991 & 0.9520 \\ 
  0.3 & 0.0100 & 0.9358 & 0.9351 & 0.9989 & 0.8378 \\ 
  0.3 & 0.0010 & 0.6700 & 0.6700 & 0.9976 & 0.5397 \\ 
  0.3 & 0.0001 & 0.3298 & 0.3298 & 0.9970 & 0.2532 \\ \hline
  0.6 & 0.0500 & 0.9079 & 0.9055 & 0.9810 & 0.7500 \\ 
  0.6 & 0.0100 & 0.7378 & 0.7375 & 0.9768 & 0.5597 \\ 
  0.6 & 0.0010 & 0.4290 & 0.4290 & 0.9720 & 0.3005 \\ 
  0.6 & 0.0001 & 0.1870 & 0.1870 & 0.9671 & 0.1301 \\ \hline
  0.9 & 0.0500 & 0.7477 & 0.7477 & 0.9441 & 0.4037 \\ 
  0.9 & 0.0100 & 0.5192 & 0.5192 & 0.9362 & 0.2438 \\ 
  0.9 & 0.0010 & 0.2400 & 0.2400 & 0.9285 & 0.0976 \\ 
  0.9 & 0.0001 & 0.0880 & 0.0880 & 0.9214 & 0.0358 \\ \hline
   \hline
\end{tabular}
\vspace{1cm}
\caption{Power performance based on 10,000 replications, $d=100$}
\end{table}

\subsection{A comprehensive comparison of CCT vs. TCCT}
In Table 2, we can see TCCT has only slightly improved test power compared to CCT. However, when some of the p-values from individual tests are close to 1, the overall test power from CCT may be significantly reduced. To show this, we modify the example given by \citet{zhongxue22} and combine several independent one-sided tests. In theory, it is possible that the test power is less than type I error $\alpha$ when testing a composite null hypothesis from one-sided test, and in such case p-value will be close to 1. We select 100 different values of $\mu$ where $\mu$ ranges from $[-c, c]$, and for each value of $\mu$, we simulate 100 normally distributed samples with mean $\mu$, and test $H_0: \mu<=0$ against $H_{\alpha}: \mu>0$. Clearly, the p-values will be close to 0 when $\mu>0$, but close to 1 when $\mu<0$. We then combine the p-values from these 100 tests by performing both CCT and TCCT. We choose different values of $c$ from 0 to 0.45, and for each $c$ we estimate the test power of both tests based on 10,000 replications.

We show the power curves of both tests in Figure 1, where the red curve represents TCCT, and the black curve is CCT. Figure 1 shows that the proposed TCCT has consistently higher power, and test power goes to 1 when $c$ value increases. However, CCT power cannot go higher than 0.5 regardless the $c$ value. This is because there are about 50\% of the combined tests with p-values close to 0 and the other 50\% close to 1, so the overall p-value will be randomly dominated by either 0 or 1, and the overall p-value is close to 1 for approximately half of the replications. This example clearly illustrates the limitation of CCT, while the proposed TCCT remains valid.

\begin{figure}
 \centerline{\includegraphics[scale=0.6]{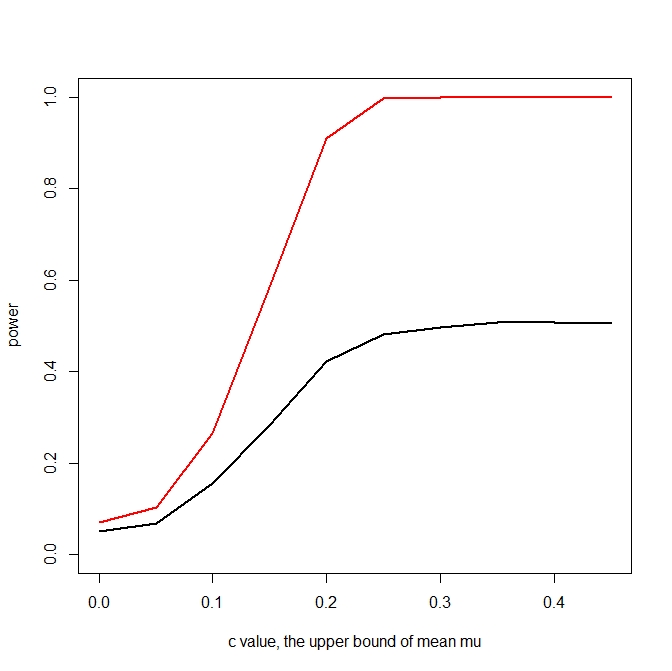}}
\caption{Test powers combining 100 one-sided tests. Red curve represents truncated Cauchy combination test (TCCT); black curve represents Cauchy combination test (CCT). $c$ value, representing the upper bound of $\mu \in [-c, c]$, ranges from 0 to 0.45.}
\end{figure}

We next give a more comprehensive comparison of TCCT versus CCT, when the p-values being combined have different patterns. We simulate 100 p-values directly from Beta($\alpha$,$\beta$) distribution. Note that Beta(1,1) is simply Unifrom(0,1) distribution, which is the null distribution of the p-values. For all parameter values of $\alpha$ and $\beta$ of the Beta distribution from 0 to 2, we compare the power performances of both combined tests and compute the power gain from our Truncated Cauchy test. We show this by using heat plots in Figure 2.

\begin{figure}
 \centerline{\includegraphics[scale=0.25]{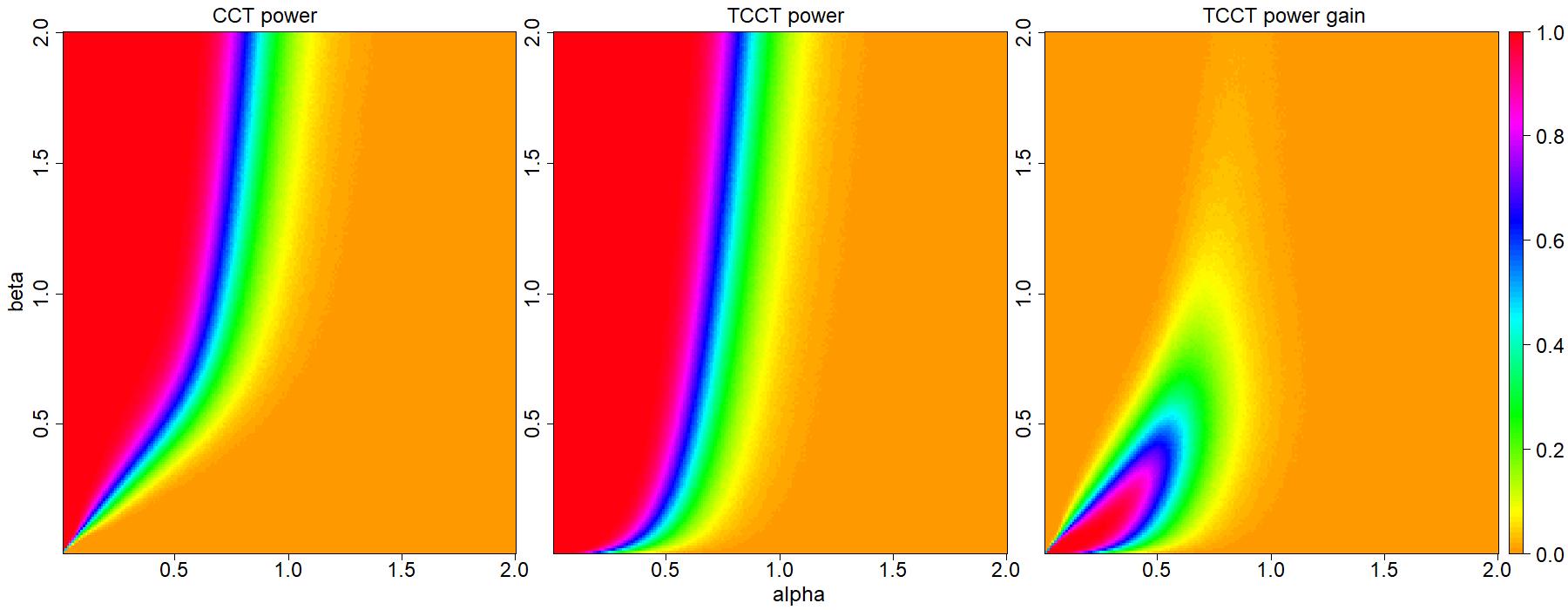}}
\caption{Test powers combining 100 p-values following Beta($\alpha$,$\beta$) distribution and power gain from truncated Cauchy combination test (TCCT) when $\alpha \in (0,2)$ and $\beta \in (0,2)$.}
\end{figure}

Figure 2 shows the power gain from TCCT can be even close to 1. This happens when both $\alpha$ and $\beta$ are small. For example, Beta(0.2, 0.1) represents the case that most p-values are concentrated around 0 and 1, but more p-values are close to 1 than close to 0 among the 100 tests. In such cases, CCT may have little power, but TCCT is not affected by those large p-values and can have the test power as large as 1.

\section{Real data applications}
\subsection{GWAS application}

Firstly, we apply TCCT to a genome-wide association study (GWAS) of 16,470 single nucleotide polymorphisms (SNPs) on 22 chromosomes. The data is publicly available from R package qqman \citep{turner18}. Different number of SNPs are collected from each chromosome, with a minimal of 535 SNPs from chromosome 22 and a maximal 1500 SNPs from chromosome 1. For each SNP, p-value of the association analysis using PLINK is reported. Within each chromosome, we apply TCCT to get an overall p-value combining all SNPs. For comparison, we also use CCT to combine those p-values. Results of both tests are summarized in Table 3.

\begin{table}[ht]
\centering
\begin{tabular}{r|rr}
  \hline
Chromosome & CCT & TCCT \\ 
  \hline
1 & 0.144 & 0.080 \\ 
  2 & 0.814 & 0.113 \\ 
  3 & 1.51E-06 & 1.51E-06 \\ 
  4 & 0.670 & 0.121 \\ 
  5 & 0.303 & 0.118 \\ 
  6 & 0.639 & 0.125 \\ 
  7 & 0.341 & 0.100 \\ 
  8 & 0.200 & 0.113 \\ 
  9 & 0.767 & 0.139 \\ 
  10 & 0.842 & 0.156 \\ 
  11 & 0.181 & 0.083 \\ 
  12 & 0.946 & 0.124 \\ 
  13 & 0.698 & 0.123 \\ 
  14 & 0.044 & 0.026 \\ 
  15 & 0.795 & 0.149 \\ 
  16 & 0.264 & 0.142 \\ 
  17 & 0.651 & 0.185 \\ 
  18 & 0.016 & 0.014 \\ 
  19 & 0.470 & 0.103 \\ 
  20 & 0.373 & 0.114 \\ 
  21 & 0.118 & 0.079 \\ 
  22 & 0.723 & 0.168 \\ 
   \hline
\end{tabular}
\vspace{1cm}
\caption{Global p-values for combining all SNPs from each of the 22 chromosomes using Cauchy combination test (CCT) and truncated Cauchy combination test (TCCT)}
\end{table}

Table 3 shows that TCCT provides considerably smaller p-values from most chromosomes than CCT, e.g., 0.124 vs. 0.946 from Chromosome 12. When very strong associations are detected (e.g., chromosome 3), both tests provide significant p-values at the same magnitude (1.51E-06). As expected from the theory, all p-values from TCCT are smaller than or equal to CCT.



\subsection{Application to longitudinal microbiome sequencing data}
We next perform another application study with longitudinal Microbiome sequencing data. Our data is collected from a gastroesophageal reflux disease (GERD) study \citep{schneeberger22}, where the primary interest is to find if the binary GERD status is associated with Microbiome, Cytokine and Bile acid data. Microbiome sequncing data are summarized by the relative abundance of operational taxonomic units (OTUs). A large portion of the relative abundance data may be 0 \citep{kaul17}, and there may be correlations between each pair of OTUs \citep{mandal15}. Bile acid data also contains a lot of zeros, and Cytokine data is approximately normally distributed.

Microbiome, Cytokine and Bile acid data are observed at 4 different timepoints. As we introduced in Section 1, both TCCT and CCT can be readily applied to longitudinal data by simply combining the p-values from each timepoint, regardless of the correlation between timepoints. Besides, the Microbiome sequencing data is grouped by 20 OTUs which are correlated with each other. In this example, we apply univariate tests for each OTU and use TCCT or CCT to combine the univariate test p-values.

To deal with the zero-inflated Microbiome and Bile Acid data, one solution is to choose a two-part model \citep{chen20}. The first part is to test the association  between GERD status and the prevalence/absence (non-zero/zero) of the data for each sample. If the data is non-zero, the second part is to test the association between GERD status and those non-zero data. We run logistic regression for the first part, and linear regression for the second part. Clearly these two parts may also be correlated. However, we can still combine both p-values by TCCT and CCT. In summary, we use TCCT and CCT to combine p-values from 2 parts, 4 time points and 20 OTUs. However, the problem of the two-part model is that for some of the OTUs, most or even all data can be non-zero, in which case the logistic regression from prevalence/absence data will yield p-value close to 1 as almost all the responses are prevalence. As we showed by simulation studies in Section 3, this may cause a problem with CCT. Therefore, we also include the classic one-part linear regression model as a comparison, although clearly the zero-inflated data are not normally distributed.  All test results are summarized in Table 4.

\begin{table}[ht]
\centering
\begin{tabular}{r|rr}
  \hline
 & CCT & TCCT \\ 
  \hline
  Cytokine & 0.036 & 0.034 \\ 
  Bile acid & 0.428 & 0.351 \\ 
  Bile acid 2 parts & 1.000 & 0.421 \\ 
  Microbiome & 0.266 & 0.162 \\ 
  Microbiome 2 parts & 1.000 & 0.267 \\ 

   \hline
\end{tabular}
\vspace{1cm}
\caption{Testing the association of Microbiome sequencing data with GERD status, individual tests are combined by Cauchy combination test (CCT) and truncated Cauchy combination test (TCCT).}
\end{table}

Table 4 shows that the TCCT p-values are consistently more significant than CCT. Besides, for the 2 parts tests where some p-values may be equal to 1 from the individual tests, CCT fails to work at all, but TCCT can still provide valid p-values. We conclude that TCCT is more robust and powerful than CCT, and it can be used conveniently for testing longitudinal data.

\section{Discussion}
In this paper, we propose a truncated Cauchy combination test (TCCT) as a p-value combination method. It needs to be noted that although related, p-value combination is a different topic than multiple testing. For p-value combination test, we only test the overall null hypothesis with one combined p-value, so there is no multiple testing issue involved, and our method is not directly compatible with multiple testing methods, such as Bofferoni correction, False discovery rate (FDR) and Family-wise error rate (FWER) \citep{dudoit08}. However, the global testing is usually the first step if multiple testing for each single hypothesis is desired. If the global null hypothesis is rejected, multiple testing approaches can next be applied to detect which single hypotheses lead to the overall rejection. Multiple testing adjustments with correlated tests is beyond the scope of this paper and needs further investigation.

We have shown that TCCT works well for longitudinal data observed at discrete timepoints. Although this should be sufficient for most application studies with longitudinal data, for some other studies, the longitudinal data are considered as a function of time $t$ which is continuous. Suppose the entire functional samples $\boldsymbol{y}(t)$ are observed during time interval $[0,T]$. For any $t \in [0,T]$, we can still compute $\boldsymbol{y}_t$ and fit it with ordinary linear regression model to achieve the p-value $p_t$. The difference from discrete case is that we can achieve uncountably many $p_t$ within time interval $[0,T]$, and we denote the p-value function as $p(t)$. In such case, it is desired to develop the functional version of TCCT, where the test statistic is
$$T_f=\int_{t=0}^T w(t) \tan{[0.5-p(t)]\pi} I(p(t)<0.5)dt.$$ 
The distribution of $T_f$ is unclear, as direct application of Theorem 1 and 2 is not possible. We aim to extend our proposed test for functional longitudinal data.


\backmatter





%

\bibliography{mybib}
\bibliographystyle{biom}






\section*{Supporting Information}
Web Appendix A, Web Appendix B and Web Appendix C referenced in Section 2 are available with this paper at the Biometrics website on Wiley Online Library. \vspace*{-8pt}





\label{lastpage}

\end{document}


\maketitle

\section*{Web Appendix A: Proofs of Lemma and Theorems}

\subsection*{Proof of Lemma 1:}
\begin{proof}
First,
$$\{f(X_i)>t \} \implies \{h(X_i)I(h(X_i)>0)>t \} \implies \{h(X_i)>t>0 \}.$$
Next,
$$\{h(X_i)>t \} \implies \{ I(h(X_i)>0)=1 \} \implies \{f(X_i)=h(X_i)I(h(X_i)>0)>t \}.$$
Therefore, $\{f(X_i)>t \} \iff \{h(X_i)>t \}$, and thus  $P(f(X_i)>t)=P(h(X_i)>t)$.
\end{proof}

\subsection*{Proof of Theorem 1:}
\begin{proof}
Let $A_{i,t}=\{f(X_i)>(1+\delta_t)t/w_i, T(\boldsymbol{X})>t \}$ and $A_t=\bigcup_{i=1}^d A_{i,t}$. Let $B_{i,t}=\{f(X_i) \leq (1+\delta_t)t/w_i, T(\boldsymbol{X})>t \}$ and $B_t=\bigcap_{i=1}^d B_{i,t}$, where constant $\delta_t$ only depends on $t$ and satisfies $\delta_t>0$, $\delta_t \to 0$ and $\delta_t t \to +\infty$ as $t \to +\infty$. Then $A_t$ and $B_t$ are disjoint, and
$$P(T(\boldsymbol{X})>t)=P(A_t)+P(B_t).$$

$\{T(\boldsymbol{X})>t \}$ implies there exists at least one $i$ such that $f(X_i)>t$, so
\begin{eqnarray*} 
P(B_t) &\leq& \sum_{i=1}^d P(B_{i,t} \cap \{ f(X_i)>t \})= \sum_{i=1}^d P(t<f(X_i) \leq (1+\delta_t)t/w_i, T(\boldsymbol{X})>t) \\
&\leq& \sum_{i=1}^d P(t/w_i d<f(X_i) \leq (1-\delta_t)t/w_i, T(\boldsymbol{X})>t)+\sum_{i=1}^d P((1-\delta_t)t/w_i<f(X_i) \leq (1+\delta_t)t/w_i) \\
&\leq& \sum_{i=1}^d P(t/w_i d<f(X_i) \leq (1-\delta_t)t/w_i, \sum_{j \neq i}^d f(X_j)>\delta_t t/w_i) \\
&& +\sum_{i=1}^d P((1-\delta_t)t/w_i<f(X_i) \leq (1+\delta_t)t/w_i) \\
&=& \sum_{i=1}^d P(t/w_i d<h(X_i) \leq (1-\delta_t)t/w_i, \sum_{j \neq i}^d h(X_j)>\delta_t t/w_i) \\
&&+\sum_{i=1}^d P((1-\delta_t)t/w_i<h(X_i) \leq (1+\delta_t)t/w_i)
\end{eqnarray*}
where the last step of equality follows Lemma 1. Denoting the first and second part of the summation by $I_1$ and $I_2$, Liu \& Xie (2020) showed that $I_1=o(\frac{1}{t})$ and $I_2=o(\frac{1}{t})$. Therefore, $P(B_t)=o(\frac{1}{t})$.

Next, we show $P(A_t)=\frac{1}{t\pi}+o(\frac{1}{t})$. By Bonferroni inequality,
$$\sum_{i=1}^d P(A_{i,t})-\sum_{1 \leq i<j \leq d} P(A_{i,t} \cap A_{j,t}) \leq P(A_t) \leq \sum_{i=1}^d P(A_{i,t}).$$
Note that $f(X_i) \geq 0$ for all $i$, so $f(X_i)>(1+\delta_t)t/w_i$ implies $T(\boldsymbol{X})>t$. Therefore,
$$A_{i,t}=\{f(X_i)>(1+\delta_t)t/w_i\}=\{h(X_i)>(1+\delta_t)t/w_i\}$$  
following Lemma 1 when $t>0$. We can then use the result in Liu \& Xie (2020), which shows $P(A_{i,t})=\frac{1}{td\pi}+o(\frac{1}{t})$, and $P(A_{i,t} \cap A_{j,t})=o(\frac{1}{t})$. It completes the proof that $P(A_t)=\frac{1}{t\pi}+o(\frac{1}{t})$. Combining with $P(B_t)=o(\frac{1}{t})$, we conclude that $P(T(\boldsymbol{X})>t)=\frac{1}{t\pi}+o(\frac{1}{t})$. Liu \& Xie (2020) also showed that $P(W_0>t)=\frac{1}{t\pi}+o(\frac{1}{t})$. Therefore,
$$\underset{t \to +\infty}{\lim} \frac{P(T(\boldsymbol{X})>t)}{P(W_0>t)}=1.$$

\end{proof}

\subsection*{Proof of Theorem 2:}
\begin{proof}
We use the same definition of $A_t$ and $B_t$, and still have 
$$
P(B_t) \leq \sum_{i=1}^d P(t<h(X_i) \leq (1-\delta_t)t/w_i, \sum_{j \neq i}^d h(X_j)>\delta_t t/w_i)+\sum_{i=1}^d P((1-\delta_t)t/w_i<h(X_i) \leq (1+\delta_t)t/w_i)
$$
and
$$\sum_{i=1}^d P(A_{i,t})-\sum_{1 \leq i<j \leq d} P(A_{i,t} \cap A_{j,t}) \leq P(A_t) \leq \sum_{i=1}^d P(A_{i,t}).$$
where $A_{i,t}=\{h(X_i)>(1+\delta_t)t/w_i\}$ as proved in Theorem 1. It then follows the proof in Liu \& Xie (2020) that $P(B_t)=o(\frac{1}{t})$ and $P(A_t)=\frac{1}{t\pi}+o(\frac{1}{t})$, which concludes that $P(T(\boldsymbol{X})>t)=\frac{1}{t\pi}+o(\frac{1}{t})$ and completes the proof of Theorem 2.
\end{proof}

\newpage

\section*{Web Appendix B: Theoretical property of $T_{min}$}

We introduced the test statistic $T_{min}$ in Section 2.2 as $T_{min}=\tan\{(0.5-p_{(1)})\pi\}/d$, where $p_{(1)}=min\{p_1,...,p_d\}$. Under the same bivariate normal distribution assumption of $\boldsymbol{X}=(X_1, X_2,..., X_d)^T$, it is equivalent to write $T_{min}(\boldsymbol{X})=\max_{1 \leq i \leq d} h(X_i)/d$, where $h(x)=\tan\{[2\Phi(|x|)-3/2]\pi\}$. Then we can show that similar to $T(\boldsymbol{X})$ in Theorem 1, the distribution of $T_{min}(\boldsymbol{X})$ also has Cauchy tail probability: 
$$\underset{t \to +\infty}{\lim} \frac{P(T_{min}(\boldsymbol{X})>t)}{P(W_0>t)}=1.$$
The proof is provided below. To show this, we use Lemma 3.3 from the Supplement of Liu \& Xie (2020), which states that if $W_0$ has a standard Cauchy distribution and $X_0$ has a standard normal distribution, then
$$P(W_0>t)=P(h(X_0)>t)=\frac{1}{t\pi}+O(1/t^3).$$

\begin{proof}
Let $A_{i,t}=\{h(X_i)>td \}$ and $A_t=\bigcup_{i=1}^d A_{i,t}$. $\{T_{min}(\boldsymbol{X})>t \}$ implies there exists at least one $i$ such that $h(X_i)>td$, so $P(T_{min}(\boldsymbol{X})>t)=P(A_t)$. Next,
$$\sum_{i=1}^d P(A_{i,t})-\sum_{1 \leq i<j \leq d} P(A_{i,t} \cap A_{j,t}) \leq P(A_t) \leq \sum_{i=1}^d P(A_{i,t}).$$
Following Lemma 3.3, 
$$P(A_{i,t})=\frac{1}{td\pi}+o(\frac{1}{t}),$$ 
so
$$\sum_{i=1}^d P(A_{i,t})=\frac{1}{t\pi}+o(\frac{1}{t}).$$
Next we show
$$\sum_{1 \leq i<j \leq d} P(A_{i,t} \cap A_{j,t})=o(\frac{1}{t}).$$
It is straightforward to show this when $X_i$ and $X_j$ are independent, because
$$P(A_{i,t} \cap A_{j,t})=P(A_{i,t})P(A_{j,t})=\frac{1}{t^2d^2\pi^2}+o(\frac{1}{t})=o(\frac{1}{t}).$$
When $X_i$ and $X_j$ are not independent, we can define $Z_{ij}=X_j-\sigma_{ij}X_i$, where $\sigma_{ij}=Cov(X_i, X_j)$. Because of the pairwise normality assumption, $Z_{ij}$ also follows normal distribution, and it is straightforward to calculate  $E(Z_{ij})=0$ and $Var(Z_{ij})=1-\sigma_{ij}^2$. Next, 
$$Cov(Z_{ij}, X_i)=Cov(X_j-\sigma_{ij}X_i, X_i)=\sigma_{ij}-\sigma_{ij}=0,$$
so $Z_{ij}$ and $X_i$ are independent. Because 
$$P(A_{i,t} \cap A_{j,t})=P(A_{i,t})P(A_{j,t}|A_{i,t}),$$ 
we next show 
$$\lim_{t \to +\infty} P(A_{j,t}|A_{i,t})=0,$$ 
and thus
$$P(A_{i,t} \cap A_{j,t})=o(\frac{1}{t}).$$
As $t \to +\infty$, it suffices to only consider the situation when $t>0$. Because $h(x)$ is a monotone function when $x>0$, we can define its inverse function as $h^{-1}$, and $A_{i,t}=\{X_i>h^{-1}(td) \}$ when $t>0$. So we can write
$$P(A_{j,t}|A_{i,t})=P(\sigma_{ij}X_i+Z_{ij}>c_t|X_i>c_t),$$
where $c_t=h^{-1}(td) \to +\infty$ as $t \to +\infty$. Let $\delta_t$ be some constant which only depends on $t$ and satisfies $\delta_t>0$, $\delta_t \to 0$ and $\delta_t c_t \to +\infty$ as $t \to +\infty$. We can further decompose $P(A_{j,t}|A_{i,t})$ as
$$P(A_{j,t}|A_{i,t})=P(\sigma_{ij}X_i+Z_{ij}>c_t, Z_{ij} \leq \delta_t c_t|X_i>c_t)+P(\sigma_{ij}X_i+Z_{ij}>c_t,Z_{ij} > \delta_t c_t|X_i>c_t)$$
It is easy to see the second part,
$$P(\sigma_{ij}X_i+Z_{ij}>c_t,Z_{ij} > \delta_t c_t|X_i>c_t) \leq P(Z_{ij} > \delta_t c_t|X_i>c_t)=P(Z_{ij} > \delta_t c_t) \to 0$$
as $t \to +\infty$, because $Z_{ij}$ and $X_i$ are independent. For the first part, 
$$P(\sigma_{ij}X_i+Z_{ij}>c_t, Z_{ij} \leq \delta_t c_t|X_i>c_t) \leq P(\sigma_{ij}X_i>(1-\delta_t)c_t|X_i>c_t).$$
Because $0<\sigma_{ij}<1$ is fixed, for large $t$, we have $1-\delta_t \to 1>\sigma_{ij}$. Then
$$P(\sigma_{ij}X_i>(1-\delta_t)c_t|X_i>c_t)=\frac{P(\sigma_{ij}X_i>(1-\delta_t)c_t)}{P(X_i>c_t)}.$$
Next we show the tail probability of a standard normal distribution, $P(X>c_t)$, can be approximated as
\begin{equation}
P(X>c_t)=\frac{1}{\sqrt{2\pi}} e^{-c_t^2/2}(\frac{1}{c_t}+o(\frac{1}{c_t}))
\end{equation}

We first obtain an upper bound for $P(X>c_t)$. For $x$ in $[c_t, +\infty)$, we have
$$P(X>c_t)=\frac{1}{\sqrt{2\pi}} \int_{c_t}^{+\infty} e^{-x^2/2} dx \leq \frac{1}{\sqrt{2\pi}} \int_{c_t}^{+\infty} \frac{x}{c_t}e^{-x^2/2} dx = \frac{1}{\sqrt{2\pi}}\frac{1}{c_t}e^{-c_t^2/2}.$$
For the lower bound, we define
$$f(c_t)=P(X>c_t)-\frac{1}{\sqrt{2\pi}}\frac{c_t}{c_t^2+1}e^{-c_t^2/2}$$
and show $f(c_t) \geq 0$ for all $c_t>0$. This is because $f(0)>0$, $f(c_t) \to 0$ as $c_t \to +\infty$, and
$$f'(c_t)=-\frac{2}{(c_t^2+1)^2}e^{-c_t^2/2}<0,$$
which implies $f(c_t)$ is strictly decreasing to 0 as $c_t \to +\infty$. Therefore,  
$$P(X>c_t) \geq \frac{1}{\sqrt{2\pi}}\frac{c_t}{c_t^2+1}e^{-c_t^2/2}.$$
Combining the upper and lower bound of $P(X>c_t)$ and considering
$$\frac{1}{c_t}-\frac{c_t}{c_t^2+1}=\frac{1}{c_t(c_t^2+1)}=o(\frac{1}{c_t}),$$
we complete the proof of Equation 1.

Following Equation 1, we have

$$\lim_{c_t \to +\infty}\frac{P(\sigma_{ij}X_i>(1-\delta_t)c_t)}{P(X_i>c_t)}=\frac{1-\delta_t}{\sigma_{ij}}\lim_{c_t \to +\infty} \exp(\frac{-(1-\delta_t)^2c_t^2}{2\sigma_{ij}^2}+\frac{c_t^2}{2})=0,$$
because
$$\frac{-(1-\delta_t)^2c_t^2}{2\sigma_{ij}^2}+\frac{c_t^2}{2} \to -\infty.$$
Therefore, we conclude that 
$$P(A_{j,t}|A_{i,t})=P(\sigma_{ij}X_i+Z_{ij}>c_t, Z_{ij} \leq \delta_t c_t|X_i>c_t) \to 0,$$ 
and thus $P(A_{j,t}|A_{i,t}) \to 0$ as $t \to +\infty$, which completes the proof of
$$P(A_{i,t} \cap A_{j,t})=o(\frac{1}{t}).$$
Because $d$ is fixed, we also have 
$$\sum_{1 \leq i<j \leq d}P(A_{i,t} \cap A_{j,t})=o(\frac{1}{t}).$$
As we have shown
$$\sum_{i=1}^d P(A_{i,t})=\frac{1}{t\pi}+o(\frac{1}{t}),$$
we finally conclude that
$$P(T_{min}(\boldsymbol{X})>t)=P(A_t)=\frac{1}{t\pi}+o(\frac{1}{t}).$$
It follows Lemma 3.3 that
$$P(W_0>t)=\frac{1}{t\pi}+o(\frac{1}{t}).$$
Therefore,
$$\underset{t \to +\infty}{\lim} \frac{P(T_{min}(\boldsymbol{X})>t)}{P(W_0>t)}=1.$$

\end{proof}

\newpage

\section*{Web Appendix C: Simulation results comparing $T_{min}$ with Tippett's test}
We have shown in Table 1 and 2 that Tippett's minimal p-value test is conservative and less powerful than CCT and TCCT. As $T_{min}$ is also transformed from the minimal p-value, we use simulation to compare its type I error and power with Tippett's minimal p-value test. We use the same parameter settings as Table 1 and 2, and show the test of $T_{min}$ has very close performance to Tippett's test. Therefore, we conclude that despite of its theoretical interest, $T_{min}$ can be less appealing in practice. Simulation results are summarized in Table A1.

\begin{table}[ht]
\centering
\begin{tabular}{rr|rr|rr}
  \hline
& & \multicolumn{2}{c|}{Type I error} & \multicolumn{2}{c}{Power} \\
$\rho$ & $\alpha$ & $T_{min}$ & Tippett & $T_{min}$ & Tippett \\ 
  \hline
0.0 & 0.0500 & 0.0489 & 0.0500 & 1.0000 & 1.0000 \\ 
  0.0 & 0.0100 & 0.0086 & 0.0086 & 0.9988 & 0.9988 \\ 
  0.0 & 0.0010 & 0.0008 & 0.0008 & 0.8656 & 0.8658 \\ 
  0.0 & 0.0001 & 0.0001 & 0.0001 & 0.4150 & 0.4150 \\ \hline
  0.3 & 0.0500 & 0.0437 & 0.0448 & 0.9515 & 0.9520 \\ 
  0.3 & 0.0100 & 0.0103 & 0.0104 & 0.8377 & 0.8378 \\ 
  0.3 & 0.0010 & 0.0010 & 0.0010 & 0.5397 & 0.5397 \\ 
  0.3 & 0.0001 & 0.0001 & 0.0001 & 0.2532 & 0.2532 \\ \hline
  0.6 & 0.0500 & 0.0264 & 0.0272 & 0.7480 & 0.7500 \\ 
  0.6 & 0.0100 & 0.0073 & 0.0073 & 0.5589 & 0.5597 \\ 
  0.6 & 0.0010 & 0.0006 & 0.0006 & 0.3004 & 0.3005 \\ 
  0.6 & 0.0001 & 0.0001 & 0.0001 & 0.1301 & 0.1301 \\ \hline
  0.9 & 0.0500 & 0.0060 & 0.0061 & 0.4016 & 0.4037 \\ 
  0.9 & 0.0100 & 0.0012 & 0.0012 & 0.2435 & 0.2438 \\ 
  0.9 & 0.0010 & 0.0002 & 0.0002 & 0.0976 & 0.0976 \\ 
  0.9 & 0.0001 & 0.0000 & 0.0000 & 0.0358 & 0.0358 \\ \hline
   \hline
\end{tabular}
\vspace{1cm}
\caption{Comparing type I error and power of $T_{min}$ and Tippett's minimal p-value test based on 10,000 replications}
\end{table}
